\documentclass[aps,prd,floatfix,superscriptaddress,nofootinbib]{revtex4}
\usepackage[latin1]{inputenc}  
\usepackage{bbm,slashed}
\usepackage{longtable}
\usepackage{mathrsfs}
\usepackage{graphicx,epsfig}
\graphicspath{{plots/}}
\usepackage{amsmath,amsfonts,amssymb}
\usepackage{booktabs}

\usepackage{epstopdf}  
   
\usepackage{url}
\usepackage{dsfont} 
\usepackage{bm,bbm,xcolor}

\usepackage{color}

\usepackage{ucs,listings}
\lstdefinelanguage{Python}{%
sensitive=true, 
}
\def\s0#1#2{\mbox{\small{$ \frac{#1}{#2} $}}}  
 
\newcommand{\N}{\mathcal{N}}    
\newcommand{\cD}{\mathcal{D}}
\newcommand{\vect}[1]{{\bm{#1}}}
\newcommand{\mbar}{{m}}

\newcommand{\abs}[1]{\left| #1 \right|}

\DeclareMathOperator{\Tr}{Tr}

 \DeclareMathAlphabet{\boldmathe}{T1}{cmr}{bx}{it}
 
 \newcommand{\FlowPy}{{FlowPy}}

\begin{document}
\title{FlowPy -- a numerical solver for functional renormalization group equations}
\author{Thomas Fischbacher}
\email{t.fischbacher@soton.ac.uk}
\affiliation{University of Southampton, School of Engineering
Sciences, Highfield Campus, University Road, SO17 1BJ Southampton,
United Kingdom}
\author{Franziska Synatschke-Czerwonka}
\affiliation{ Theoretisch-Physikalisches Institut, Friedrich-Schiller-Universit{\"a}t
Jena,
Max-Wien-Platz 1, D-07743 Jena, Germany}

\begin{abstract}
\noindent
FlowPy is a numerical toolbox for the solution of partial differential
equations encountered in Functional Renormalization Group
equations. This toolbox compiles flow equations to fast machine code
and is able to handle coupled systems of flow equations with full
momentum dependence, which furthermore may be given implicitly.

\end{abstract}
\maketitle

\section{Introduction}
\noindent
In recent years the functional renormalization group (FRG) has  been
successfully applied to a wide variety of nonperturbative problems
such as critical phenomena, fermionic systems, gauge theories,
supersymmetry and quantum gravity, see
\cite{Litim:1998nf,Aoki:2000wm,Berges:2000ew,Polonyi:2001se,Pawlowski:2005xe,Gies:2006wv,Sonoda:2007av,Delamotte:2007pf,Bagnuls:2000ae,Synatschke:2009nm,Synatschke:2010ub,Pawlowski:2010ht,Braun:2011pp}
for reviews. Although these systems are very different in their
physical nature, the flow equations always have a similar structure.

The aim of the FRG is to calculate the generating functionals of 1PI
correlation functions from which the dynamics of the theory can be
inferred. The core ingredient is the scale dependent effective action
denoted by $\Gamma_k$ with the RG scale $k$. It interpolates between a
microscopic description through the classical action at some UV scale
$k=\Lambda$ and a macroscopic description at low energy scales $k=0$
through the full quantum effective action. The RG scale $k$ serves as
an infrared regulator suppressing all fluctuations with momentum
smaller than $k$. Thus, for $k=0$ all fluctuations are taken into
account and we have obtained a full solution of the quantum
theory. The flow of the scale dependent effective action is governed
by the Wetterich equation \cite{Wetterich:1992yh}
\begin{equation}
 \partial_k\Gamma_k=
 \frac12 \Tr\left[\left(\Gamma_k^{(2)}+ R_k\right)^{-1}\partial_k  R_k\right],
\label{eq:lpa1}
\end{equation}
with $\Gamma^{(2)}_k$ being the second functional derivative of the
effective action.  The momentum-dependent regulator function $R_k$ in
Eq.~(\ref{eq:lpa1}) establishes the IR suppression of modes below
$k$. In the general case, three properties of the regulator $R_k(p)$
are essential: (i) $R_k(p)|_{p^2/k^2\to 0} >0$ which implements the IR
regularization, (ii) $R_k(p)|_{k^2/p^2\to 0} =0$ which guarantees that
the regulator vanishes for $k\to0$, (iii)
$R_k(p)|_{k\to\Lambda\to\infty}\to \infty$ which serves to fix the
theory at the classical action in the UV.  Different functional forms
of $R_k$ correspond to different RG trajectories manifesting the RG
scheme dependence, but the end point $\Gamma_{k\to 0}\to \Gamma$
remains invariant.

Solving the partial nonlinear differential equation \eqref{eq:lpa1}
head on is impossible in most cases. Thus approximations for the
effective action have to be introduced resulting in a system of
coupled differential equations. Recently, a Mathematica extension
which is able to derive Dyson-Schwinger equations and functional
renormalization group equations was published
\cite{Huber:2011qr}. However, solving these systems beyond the most
simple approximations is a numerical challenge. For some systems,
e.\,g. supersymmetric quantum mechanics \cite{Synatschke:2008pv}
and the two dimensional $\mathcal N=1$ Wess-Zumino model
\cite{Synatschke:2009nm}, the coefficient function of the highest
derivative can become singular. This implies that solving the
differential equations numerically has to be done with great care. If
the equations are solved exactly, the singularity is never reached but
rather the flow is repelled if it comes close to the singularity. A
numerical solution has to take this behavior into account.

Including a full momentum dependence has become more important over
the last few years. This leads to a much higher numerical effort for
the solution of the flow equations. Full momentum dependence of
propagators and vertices has previously been treated successfully in
the literature
\cite{Ellwanger:1995qf,Pawlowski:2003hq,Fischer:2004uk,Fischer:2008uz,Blaizot:2006vr,Blaizot:2005wd,Blaizot:2005xy,Benitez:2009xg,Diehl:2007xz,Fister:2011uw}.

This article presents a numerical toolbox called \FlowPy{} for the
solution of a broad class of partial differential equations that are
encountered in the study of the Functional Renormalization Group
equations. Specifically, FlowPy supports full momentum
dependence of the flowing function, systems of coupled differential
equations, and also implicit specification of the $k$-derivative (as
they are encountered in the flow equation for a field dependent wave
function).

This paper is organized as follows: In Sec.~\ref{sec:NumericalSetup}
we describe the numerical setup for \FlowPy{}. In
Sec.~\ref{sec:UsingFlowpy} to Sec.~\ref{sec:fileformat} we describe how \FlowPy{} is used, which
parameters it takes and how they are specified. In Sec. \ref{sec:Examples}
we discuss both simple and typical examples of differential equations
in the study of the Functional Renormalization Group. We give detailed
examples for Python scripts that demonstrate how \FlowPy{} can be used
in practice. Our aim is to demonstrate how FlowPy is used, thus we
just take flow equations from the literature without any derivation.
As FlowPy is a general solver for partial differential equations we
will denote the flow functions with $f(k,x)$ most of the time, unless
we consider some special physical system.

\section{A Numerical Framework for RG Flows}
\label{sec:NumericalSetup}
\noindent
The main design choices when building a framework for renormalization
group flows are about how to support a reasonably large class of
interesting problems while keeping code complexity manageable. 

One of the most challenging aspects of functional
renormalization group problems is that the rate of change of
the flow function~$f(k,x)$ can, for each value of the scale
parameter~$k$ and at each point~$x$, receive contributions
from all other points~$x'$. Hence, we are
dealing with non-local partial integro-differential
equations, or coupled systems of such equations, which may
furthermore
contain higher derivatives with respect to the
coordinate~$x$, and
potentially be given in implicit form only.

Conceptually, a numerical approach to problems of this nature involves
numerical ODE solving (after discretization of the $x$~range, as
discussed below), numerical interpolation, numerical integration, as
well as numerical differentiation. While an advanced numerical
approach to such problems would perhaps develop a sophisticated
combined (adaptive) discretization scheme that handles these different
aspects in an unified fashion, here we contend ourselves with
combining functionality from readily available libraries
(specifically, functions from ODEPACK\cite{ODEPACK},
FITPACK\cite{fitpack}, and QUADPACK\cite{quadpack}) to solve the first
three of the aforementioned tasks. Rather than working with these
libraries directly, we use wrappers available in Scientific
Python~\cite{SciPy}, as the flexibility provided by the Python
programming language is very helpful for addressing some subtle
aspects of the task.

The {\FlowPy} package's objective is to numerically solve equations
(resp. equation systems) of this type through discretization of the
$x$-range with $x$ denoting for example a field variable or a
momentum. Choosing a number of support
points~$x_j,j\in\{1,2,\ldots,j_{\rm max}\}$, a PDE for~$f(k,x)$ of the
form
\begin{equation}
\frac{d}{dk}\;f(k,x)=G[k;x;f; \partial_x f; \partial_x^2 f;\ldots]
\end{equation}
(where $G$ is some functional) gets turned into a set of coupled ODEs
of the form $d/dk f_j(k) = G_j[k;\vec f;\vec f';\vec f'';\ldots]$ with
$f_j(k)=f(k,x_j)$ and $\vec f'$ being the vector $\partial_x f(k;x_j)$
etc.\,. If the functional~$G$ involves integration over the position
parameter (as it often does), the computational effort needed to
evaluate the right hand side once (in order to numerically integrate
the ODE) will grow quadratically with the number of support points. At
the time of the first release of {\FlowPy}, choosing the number of
support points $j_{\rm max}$ somewhere between~20 and~100 and using a
geometric distribution for the $x_j$ seems an appropriate choice for
many problems.

An earlier prototype of {\FlowPy}, which was used to do the
calculations underlying~\cite{Synatschke:2010jn}, was only concerned
with performing the ODE integrations after $x$-discretization and
required writing low level C code to specify the right
hand side of the flow equation. It was soon found that this was a fairly tedious procedure
that in particular needed some quite specific computing expertise
beyond what may reasonably be expected from physicists wanting to
solve RG flow equations. For this reason, the version of {\FlowPy}
described here was extended with an equation parser and code generator
that automatically translates flow equation specifications to machine
code and then loads this for fast execution. This `equation compiler'
is now the largest component of the {\FlowPy} package.  A similar
program package to automate the calculations of Dyson-Schwinger
equations was presented in \cite{Huber:2011xc}.

\section{Using FlowPy}
\label{sec:UsingFlowpy}
\noindent
{\FlowPy} needs the following packages to be pre-installed:
\begin{itemize}
	\item Python (2.$x$ with $x\ge 6$)
	\item Scientific Python (SciPy)
	\item The `NumPy' package for specialized numerical arrays
	\item The `python-simpleparse' parser generator package
	\item A C compiler such as gcc (the default compiler used by
          {\FlowPy}), as well as the Python library header files
          (especially \texttt{Python.h}).
\end{itemize}
If the Python extension packages are installed into a location not
normally searched by Python, the environment variable
\texttt{PYTHONPATH} must be configured to include the
installation-specific Python extension module path
(e.g. \texttt{\$HOME/lib/python2.7/site-packages}).

The source code that accompanies this article can be downloaded either
together with the arXiv preprint source from
\begin{center}
http://arxiv.org/e-print/1202.5984,
\end{center}
or through the `ancilliary files'
link on arXiv. The distribution contains the {\FlowPy} code, a html
documentation, license and installation information, and some
examples.

Python header files are required by {\FlowPy} as it will use the
C~compiler to translate user-specified equations to a compiled
Python extension module. This {\FlowPy}-generated C module refers to a
number of low-level Python definitions. On Linux systems, these header
files usually come in a package named \texttt{python2.7-dev} or
similar.

If these packages are installed and Python is configured correctly,
running \lstinline!python tests.py!  in a console in the project folder
performs a test.  If it is successful the output is similar to:

{\scriptsize
\begin{lstlisting}[frame=single]
......
-------------------------
Ran 6 tests in 26.568s

OK 	
\end{lstlisting}}%
\noindent
Users are strongly advised to perform this check in order to ensure
{\FlowPy} has been installed and set up correctly before using it.

A complete example showing basic use of {\FlowPy} is provided in the
file \texttt{demo.py}. This shows how to compute a renormalization
group flow from~$k=10^5$ to $k=10^{-3}$ with~$20$ intermediate
$k$-steps in geometric distribution for the flow equation
\begin{align}
\nonumber
\frac{d}{dk} f(k,x) =&
 \int\limits_0^kd q \int\limits_{-\pi}^{\pi}d\phi\,
  W(C_-(x^2-k^2,q,\phi),C_+(x^2-k^2,q,\phi))\\
  &\phantom{\int d q \int d\phi}\cdot(1-0.1\frac{d}{dk} f(k,x))\intertext{with}
W(a,b)=&\frac{a\cdot b}{(2\cdot10^{10}+a^2+b^2)},\\
C_-(p_1,p_2,\phi)=&p_1^2+p_2^2-2p_1p_2\cos\phi\\
C_+(p_1,p_2,\phi)=&p_1^2+p_2^2+2p_1p_2\cos\phi
 \end{align}

\noindent
The \texttt{demo.py} Python code is shown in Listing~\ref{fig:demopy}.

When defining flow problems that involve double integration, it
might make sense to check whether changing the order in which
integrations are specified (and hence performed) makes a difference
with respect to performance. Depending on the nature of the problem,
this can -- in the present version of FlowPy -- have a major impact.

{\scriptsize
\begin{lstlisting}[frame=single,caption=The \lstinline!demo.py! Python
  code,label=fig:demopy]	
# demo.py
from FlowPy import grange, flowproblem, make_flow_logger, make_lhs_iterator 


# Note: triple-quote """ bounds multi-line Python strings
eqns="""
d/dk f(k,x) =
 integral[d q from 0 to k,d phi from -pi to pi]
  W(Cminus(x^2-k^2,q,phi),Cplus(x^2-k^2,q,phi))*(1-0.1*d/dk f(k,x));
  
FLOWSTART f(k,x)=1;

# === Helper functions ===

Cminus(p1,p2,alpha) = p1^2+p2^2-1*p1*p2*cos(alpha);
Cplus(p1,p2,alpha) = p1^2+p2^2+1*p1*p2*cos(alpha);

W(a,b)=a*b/(p0^2+a^2+b^2);

# === Constants ===
pi = 3.141592653587983;
p0 = 2.0e5;
"""

xs_plus=grange(0.1,100,7) # geometrically subdivided interval
xs=[-xj for xj in xs_plus]+[0.0]+xs_plus # x-discretization

fp=flowproblem("demo_problem", # problem/project name
               xs,             # position discretization
               equations=eqns, # the equations
               # Perform 20 k-steps geometrically distributed
               # between 10^5 and 10^(-3):
               ks=grange(1e5,1e-3,20),
               # Log flow data to file:
               log_state=make_flow_logger("demo_problem.flow"),
               #
               # For implicit flow equations that involve
               # the LHS expression (here d/dk f(k,x))
               # on the RHS: how to handle iterative determination
               # of a self-consistent RHS value:
               #
               decide_iterate=make_lhs_iterator(loops=0),
               #
               # --- Discretization Parameters below ---
               #
               # Step size for higher-order numerical differentiation:
               eps_diff=1e-4, # this also is the default.
               diff_ord=4, # this also is the default.
               interpolation_kind=4, # 4th order interp (also default).
               verbose=2, # verbose=0 turns off extra flow debug messages
               )

fp.flow()

# After this, data will be in the file "demo_problem.flow".
\end{lstlisting}}

The {\FlowPy} package provides the \texttt{flowproblem} class as well
as some auxiliary functions such as \texttt{grange} (to produce a
geometric distribution of numbers in a given range) and
{\lstinline!linrange! (to produce an arithmetic distribution of
  numbers in a given range)}, \texttt{make\_flow\_logger} (to produce
an object that can be used as \texttt{log\_state} parameter to the
\texttt{flowproblem} constructor function) and
\texttt{make\_lhs\_iterator} (to produce a decision function that can
be used as \texttt{decide\_iterate} parameter). The only relevant
method of the flow problem class is the \texttt{obj.flow()} method
that executes the solution of the problem.

In order to solve the flow equation, a \lstinline!flowproblem! object
is created. The default values for parameters are given in the
following:

{\scriptsize
\begin{lstlisting}[frame=single]	
FlowPy.flowproblem(problem_name,
                 xs, 
                 equations,
                 ks=grange(1e5,1e-3,20),
                 log_state=None,
                 eps_diff=1e-4,
                 diff_ord=4,
                 interpolation_kind=4, 
                 decide_iterate=make_lhs_iterator(eps_abs=1e-8),
                 verbose=0, 
                 )
\end{lstlisting}}
\noindent
The mandatory parameters to \lstinline!FlowPy.flowproblem()! are:

\begin{raggedleft}
\begin{tabular}{@{}lp{0.75\textwidth}@{}}
	\lstinline!problem_name! & The problem name -- this is also used to 
        create a directory that will contain machine-generated low
        level code which has been produced from the user specified equations\\
        \lstinline!xs! & a list of support points for $x$-discretization\\
        \lstinline!equations! & the flow equations, as a (usually
                                long) string. These will be handed 
                                to the parser and code generator.\\
\end{tabular}
\end{raggedleft}

\noindent
Other parameters that can be used for writing a logfile, writing to
standard output in order to allow immediate supervision of the flow,
interpolation order, steps for the $k$ integration etc. are:

\begin{raggedleft}
\vspace{-.75\baselineskip}
\begin{longtable}{@{}lp{0.68\textwidth}@{}}
          \lstinline!ks! & $k$-steps for the flow\\
 	  \lstinline!log_state! & function that logs a flow state with call signature \lstinline!f(ff_names,xs,k,ff_ys,ff_ydots)!.                           Default \lstinline!None! means: Do nothing.
                           \lstinline!ff_ys! and \lstinline!ff_ydots!
                           are lists of arrays, one for each flow
                           function, containing the values at the
                           discretized support points \lstinline!xs! for
                           the given value of \texttt{k}.\\          
          \lstinline!eps_diff! & step size for numerical differentiation in right hand side expressions\\            
          \lstinline!interpolation_kind! & specifies interpolation to
			be used (as a parameter to \texttt{scipy.interpolate.interp1d})\\
          \lstinline!diff_ord! & Numerical differentiation will be done in a way that is correct to this given order\\
          \lstinline!decide_iterate! & a decision function \lstinline!f! mapping \lstinline!f(k,history)! to True / False,
                                True meaning ``do another iteration to determine LHS \lstinline!d/dk!''
                                (cf. documentation of \lstinline!make_lhs_iterator!) \\
          \lstinline!verbose! & determine verbosity level for reporting\\
          \lstinline!cc_call! & Template pattern for calling the C
          compiler.\\
          &Default: \lstinline!gcc 2>$F.cc_log -I!\\& \lstinline!/usr/include/python$V -fPIC -shared -o!\\ 
				&\lstinline!$F.so $F.c!
\end{longtable}
\vspace{-.75\baselineskip}
\end{raggedleft} 

\noindent
The flow equations are defined as a string in between delimiters
\lstinline!"""! and consist of a collection of definitions, each
ending in a semicolon~`\texttt{;}'. There are four possible kinds of
definitions:
 \begin{itemize}
 	\item Constant definitions such as \lstinline!c=100;!
 	\item Helper functions such as \lstinline!hb(vdiff,k,z)=vdiff+k*z^2;!
 	\item Starting conditions for the flow equation such as
          \lstinline!FLOWSTART f(k,x)=1;!.
	\item Flow equations such as \lstinline!d/dk f(k,x) = [rhs];! \newline 
	The first argument in the functions always has to be the RG-parameter $k$, the second can be a field variable, a momentum etc.
 \end{itemize}
There are two kinds of boundary conditions: For each flow function,
we have to specify the values at the starting value of~$k$. This is
done with a \texttt{FLOWSTART} definition. At the boundaries of the
$x$-discretization range~$x_{\rm min/max}$, the flow will, for every
flow function~$f$ and every value of~$k$, be clamped to 
\[(d/dk)
f(k,x_{\rm min/max})=0.\] It is planned to drop this restriction in a
future release of FlowPy to allow for more flexible boundary
conditions.

The order of definitions does not matter because {\FlowPy} will sort
them automatically into the right order, and hence they can be
written down in the way that best describes the problem. However,
there are some obvious constraints: Definitions of
constants and auxiliary functions may involve auxiliary functions or
other constants, but of course there must not be circular
dependencies in these definitions. The compiler will detect and
report circularity if this rule is violated. Also, while flow
functions may be implicit i.\,e. the right hand side depends on the
left hand side, and may involve auxiliary functions, flow functions
must not be used on the right hand side of the definition of an
auxiliary function or constant.

Flow equations can be coupled, given implicitly, can contain at most
two integrations and can contain derivatives (also of higher order)
with respect to the second argument. {\FlowPy} will report an error if
flow equations contain derivatives too high for the interpolation
scheme used. A number of basic examples are discussed in
Section~\ref{sec:Examples}.

\section{FlowPy's Internal Mechanics}
\label{sec:interna}
\noindent
In order to use {\FlowPy} to full effect, it is useful to have a basic
mental model of its internal design.
When a \texttt{flowproblem} object is created, this
will use {\FlowPy}'s built-in compiler to translate the user-specified
flow problem equations to C code that subsequently gets compiled
(calling the external C compiler) to a Python extension module. The C
code, as well as the Python module object code and a compilation logfile
will be placed in a special directory that is created with the name of
the user-provided flow problem. These low-level files are named
\texttt{flowpycoreN.*}, where \texttt{N} starts at~1 and is increased
by one for every new problem produced by the same Python
process. (This is needed to inform the Python module import mechanism
that different \texttt{flowpycoreN.so} objects specify different
problems.) When running multiple {\FlowPy} processes simultaneously on a
cluster from within a cluster-wide visible directory, it is
\emph{strongly} advised to ensure that the \texttt{problem\_name}
parameter contains a task-id so that different tasks running on
different computer nodes use different auxiliary directories for their
low-level modules.

The grammar specifying flow problems is defined in the {\FlowPy} file
\texttt{FlowPy\_grammar.py}; the parser that produces the parse tree
from a problem specification is given in
\texttt{FlowPy\_parser.py}. The file \texttt{FlowPy\_builtins.py}
contains definitions of built-in {\FlowPy} functions (such as
\texttt{exp}, \texttt{cos}, etc.), as well as the skeleton for the C
code of the Python extension module to be generated. The generated
Python module will contain code that initializes all constants,
defines user-specified auxiliary functions, parameter-dependent
integration boundaries, initial conditions for flow functions, as well
as flow-function right-hand sides. When the evaluation of a flow
function right-hand side needs to access an interpolated value of the
flow function, or any of its derivatives, the C code will perform a
callback into Python, evaluating a generic interpolating function that
was provided to it by {\FlowPy}. A considerable amount of code magic
is hidden in these Python-defined interpolating functions that also
can provide interpolated values for derivatives. Typically, these are
complex closures involving various SciPy functions that interface
FORTRAN code.

\section{Format of Output Files}
\label{sec:fileformat}
\noindent
In Listing~\ref{fig:SampleOutput} we show as an example part of a logfile from Sec.~\ref{logfile} provided by FlowPy.
\begin{center}
{\scriptsize
\begin{lstlisting}[frame=single,caption=Sample output provided by FlowPy,label=fig:SampleOutput]
# === FLOWPY1.0 LOG FILE ===
# (Please remember to cite the FlowPy article in your research!) 
# xs=[-1          , -0.909091   , +0          , +0.909091   , +1]
# flowfuns=['E', 'lambda', 'omega']
# k ; nr_flowfun ; ys/ydots=0/1 ; ys/ydots[0] ; ys/ydots[1] ; ...
+100000   0  0	+0  +0            +0          	+0          	+0
+100000   1  0	+1  +1            +1          	+1          	+1
+100000   2  0	+1  +1            +1          	+1          	+1
+100000   0  1	+0  -3.1831e-11   -3.1831e-11 	-3.1831e-11 	+0
+100000   1  1	+0  +1.90986e-20  +1.90986e-20	+1.90986e-20	+0
+100000   2  1	+0  -3.1831e-11   -3.1831e-11 	-3.1831e-11 	+0
+39810.7  0  0	+0  +4.90063e-06  +4.90063e-06	+4.90063e-06	+0
+39810.7  1  0	+1  +1            +1          	+1          	+1
+39810.7  2  0	+1  +1            +1          	+1          	+1
+39810.7  0  1	+0  -2.00841e-10  -2.00841e-10	-2.00841e-10	+0
+39810.7  1  1	+0  +7.60329e-19  +7.60329e-19	+7.60329e-19	+0
+39810.7  2  1	+0  -2.0084e-10   -2.0084e-10 	-2.0084e-10 	+0
+15848.9  0  0	+0  +1.70471e-05  +1.70471e-05	+1.70471e-05	+0
+15848.9  1  0	+1  +1            +1          	+1          	+1
+15848.9  2  0	+1  +1.00002      +1.00002    	+1.00002    	+1
+15848.9  0  1	+0  -1.26724e-09  -1.26724e-09	-1.26724e-09	+0
+15848.9  1  1	+0  +3.02692e-17  +3.02692e-17	+3.02692e-17	+0
+15848.9  2  1	+0  -1.26721e-09  -1.26721e-09	-1.26721e-09	+0
+6309.57  0  0	+0  +4.74848e-05  +4.74848e-05	+4.74848e-05	+0
+6309.57  1  0	+1  +1            +1          	+1          	+1
+6309.57  2  0	+1  +1.00005      +1.00005    	+1.00005    	+1
+6309.57  0  1	+0  -7.99596e-09  -7.99596e-09	-7.99596e-09	+0
+6309.57  1  1	+0  +1.20504e-15  +1.20504e-15	+1.20504e-15	+0
+6309.57  2  1	+0  -7.99558e-09  -7.99558e-09	-7.99558e-09	+0

...
\end{lstlisting}}
\end{center}
On the third line, starting with \lstinline!# xs! are the
discretization points, the fourth line gives the functions for which
the flow equations were solved. Starting from line six, the results
from the solution of the flow equation are listed. The first column
denotes the $k$ values, the second column is a number corresponding to
the respective function, \lstinline!E!, \lstinline!lambda! and
\lstinline!omega! in the logfile above. \lstinline!0! and
\lstinline!1! in the third column stand for the function and its
derivative respectively. The following columns display the values of
the functions evaluated at the discretization points. As can be seen,
the values at the boundary are fixed throughout the flow.

Note that the logfile will not be erased if a new run of FlowPy 
is started for the same flowproblem. The entries for the new logfile are written below the old one.

\section{Examples}
\label{sec:Examples}
\noindent
In this section we will show how typical examples for flow equations
are solved with \FlowPy{} and the results are compared with solutions
obtained from \textsc{Mathematica} and SciPy, in order to establish
that FlowPy solves these problems correctly. In the subsequent
sections we will also display examples of Python scripts which
demonstrate the specification of the flow equations and the parameters
that \FlowPy{} takes.

\subsection{Simple examples}
\noindent
The first five examples are devoted to the solution of simple
differential equations, most of which have an analytic solution. The
purpose of these first examples, which are contained in the
\texttt{tests.py} test file, is to demonstrate in a readily verifiable
way that {\FlowPy} correctly works as claimed.
\subsubsection{Constant growth}
\noindent
As a first example we solve the differential equation
\begin{align}
		\frac{\partial}{\partial k}f(k,x)=-1
\end{align}
describing a constant growth with the flow parameter $k$ lying between $k\in[\Lambda,k_0]$ with $\Lambda=110$.  The function $f(k,x)$ is specified at the scale $\Lambda$ and the boundary conditions for $x_{\rm{start}}=0$ and $x_{\rm{end}}=10$ are chosen in the following way at $k=\Lambda$ and $x=x_{\rm start/end}$:
\begin{align}
f(\Lambda,x)=x,\quad f(k,x_{\rm{start}})=f(\Lambda,x_{\rm{start}}),\quad f(k,x_{\rm{end}})=f(\Lambda,x_{\rm{end}})
\end{align}
The analytic solution of this equation with the above starting condition is
\begin{align}
	f(k,x)=\Lambda-k+x=110-k+x.\label{eq:Solution1}
\end{align}
The corresponding {\FlowPy} problem specification is (note that 
$d/dk f$ is specified as being negative as we are flowing from large
to small~$k$):

{\scriptsize
\begin{lstlisting}[frame=single]
flowproblem("example1",
            xs=[float(n) for n in xrange(11)],
            equations="""
            d/dk f(k,x)=-1;
            FLOWSTART f(k,x)=x;
            """,
            log_state=logger1,
            ks=grange(110,10,1),
            decide_iterate=make_lhs_iterator(loops=0))
\end{lstlisting}}
\noindent
Solving this flow equation with \FlowPy{} yields the correct solution
as can be easily checked with eq.~\eqref{eq:Solution1}:

\begin{center}
\begin{tabular}{c|ccccccccccc}
	$x$     	&0&1          	&2          	&3          	&4          	&5          	&6          	&7          	&8          	&9          	&10     \\   \hline
	$f(k=10,x)$&0&101        	&102        	&103        	&104        	&105        	&106        	&107        	&108        	&109        	 &10
\end{tabular}
\end{center}
As expected the function at the boundary is fixed to the values at
$k=\Lambda$ throughout the flow.

\subsubsection{Exponential growth}
\noindent
The second example is a differential equation describing
exponential growth,
\begin{align}
	  \frac{\partial}{\partial k}f(k,x) = -0.01\cdot f(k,x),
\end{align}
with $\Lambda=110$, $x_{\rm{start}}=0$ and $x_{\rm{end}}=10$ and the values at the boundary fixed to the value at $k=\Lambda$ for all values of $k$, 
\begin{align}
f(\Lambda,x)=x,\quad f(k,x_{\rm{start}})=f(\Lambda,x_{\rm{start}}),\quad f(k,x_{\rm{end}})=f(\Lambda,x_{\rm{end}}).
\end{align}
The solution of this equation with the above starting condition is
\begin{align}
	f(k,x)=x e^{-0.01(k-\Lambda)}\label{eq:Solution2}
\end{align}
In FlowPy the equations are specified in the following way:
{\scriptsize
\begin{lstlisting}[frame=single]
f=flowproblem("example2",
	     xs=xs,
             equations="""
             d/dk f(k,x) = -0.01*f(k,x);
             FLOWSTART f(k,x) = x;
             """,
             log_state=logger2,
             ks=grange(110,10,1),                      
             decide_iterate=make_lhs_iterator(loops=0))
\end{lstlisting}}
\noindent
The numerical solution of this equation with \FlowPy{} yields
\begin{center}
\begin{tabular}{c|ccccccccccc}
	$x$ &0          	&1          	&2          	&3          	&4          	&5          	&6          	&7          	&8          	&9          	&10   \\      \hline
	$f(k=10,x)$&0          	&2.7    	&5.4    	&8.2    	&10.9    	&13.6    	&16.3    	&19.0     	&21.8    	&24.5    	&10  
\end{tabular}
\end{center}
which is in accordance with the analytic solution from
eq.~\eqref{eq:Solution2}.  Again, the values at the boundary do not
flow.

\subsubsection{Implicit ${\partial}/{\partial k}f(k,x)$}
\noindent
Especially when considering a problem additionally containing a wave
function renormalization, the flow equation for the wave function
renormalization is given implicitly. In order to demonstrate that
\FlowPy{} can also solve implicitly given functions, we discuss the
differential equation given by
\begin{align}
         \frac{\partial}{\partial k} f(k,x) = -1.0+0.5\frac{\partial}{\partial k} f(k,x)
\end{align}
with the starting condition $f(\Lambda,x)=x$
and $x_{\rm{start}}$, $x_{\rm{end}}$ and $\Lambda$ the same as above. 
The analytic solution to this equation is 
\begin{align}
	f(k,x)=2\Lambda-2k+x.
\end{align}
The FlowPy code reads:
{\scriptsize
\begin{lstlisting}[frame=single]
 f=flowproblem("example3",
              xs=xs,
              equations="""
              d/dk f(k,x) = -1.0+0.5*d/dk f(k,x);
              FLOWSTART f(k,x) = x;
              """,
              log_state=logger3,
              ks=grange(110,10,1),                      
              decide_iterate=make_lhs_iterator(eps_abs=1e-6))
\end{lstlisting}}
\noindent
and solving numerically with \FlowPy{} yields
\begin{center}
	\begin{tabular}{c|ccccccccccc}
		$x$&0          	&1          	&2          	&3          	&4          	&5          	&6          	&7          	&8          	&9          	&10         \\\hline
	$f(k=10,x)$ 	&0          	&201        	&202        	&203        	&204        	&205        	&206        	&207        	&208        	&209        	&10         
	\end{tabular}
\end{center}

\subsubsection{Differential equation with integral on the right hand side}
\noindent
Often in the study of the renormalization group equations, the right
hand side is given as an integral in the momentum. \FlowPy{} can also
handle  such a right hand side as is demonstrated in this section.
The flow rate is constant for each point, but for each $x$-value, we
express the flow rate as an integral:
\begin{align}
	\frac{\partial}{\partial k}f(k,x) = -\int_0^xdy\, y^2 
\end{align}
The starting condition is chosen to be $f(k=\Lambda,x)=x$
and the solution to this equation is
\begin{align}
	f(k,x)=\frac13(\Lambda-k)x^3+x.
\end{align}
The flowproblem takes the form
{\scriptsize
\begin{lstlisting}[frame=single]
fp=flowproblem("example4",
             xs=xs,
             equations="""
             d/dk f(k,x) = integral [dq from 0 to x] -q^2;
             FLOWSTART f(k,x) = x;
             """,
             log_state=logger4,
             ks=grange(110,10,1),                      
             decide_iterate=make_lhs_iterator(loops=0))
\end{lstlisting}}
\noindent
and the numerical solution with \FlowPy{} is 
\begin{center}
	\begin{tabular}{c|ccccccccccc}
$x$	&0          	&1          	&2          	&3          	&4          	&5          	&6          	&7          	&8          	&9          	&10         \\\hline
 $f(k=10,x)$	&0          	&34   	&269    	&903        	&2137    	&4171    	&7206       	&11440    	&17075    	&24309      	&10 
	\end{tabular}
\end{center}
Again, the boundary values are fixed throughout the flow.

\subsubsection{Heat equation}
\noindent
As a more complicated example we solve a heat flow problem        
        \begin{align} 	
                  \frac{\partial}{\partial k}f(k,x) = -0.01\cdot f''(k,x),\quad
                       f(k=\Lambda,x) = \exp{\left(\frac{-0.5(x-5.0)^2}{2.0^2}\right)}
        \end{align}
         with \FlowPy{}. The flowproblem reads:
         {\scriptsize\begin{lstlisting}[frame=single]
fp=flowproblem("example5",
             xs=xs,
             equations="""
             d/dk f(k,p) = -0.01*f''(k,p);
             FLOWSTART f(k,p) = exp(-0.5*(p-5.0)^2/(2.0^2));
             """,
             ks=ks,
             log_state=logger5,
             decide_iterate=make_lhs_iterator(loops=0))
         \end{lstlisting}}
\noindent       
         Values obtained with the \lstinline!odeint!
         routine from SciPy and a somewhat less sophisticated
         discretization of the second derivative are given in the table below.        
        \begin{center}
        	\begin{tabular}{c|c|c|c}
		$x_s$ & Solution from SciPy & Solution from FlowPy&Deviation in \% \\ \hline
0	&0.04394	&0.04394	&0.000\\
0.1	&0.05730	&0.05731	&0.006\\
0.2	&0.07068	&0.07069	&0.010\\
0.3	&0.08410	&0.08411	&0.012\\
0.4	&0.09756	&0.09757	&0.013\\
0.5	&0.11109	&0.11110	&0.014\\
0.6	&0.12468	&0.12470	&0.015\\
0.7	&0.13837	&0.13839	&0.014\\
0.8	&0.15215	&0.15217	&0.013\\
0.9	&0.16603	&0.16605	&0.013\\
1	&0.18001	&0.18003	&0.012\\
1.1	&0.19410	&0.19412	&0.011\\
1.2	&0.20829	&0.20831	&0.010\\
1.3	&0.22258	&0.22260	&0.009\\
1.4	&0.23697	&0.23699	&0.008\\
1.5	&0.25144	&0.25146	&0.006\\
1.6	&0.26599	&0.26600	&0.005\\
1.7	&0.2806		&0.28061	&0.004\\
1.8	&0.29526	&0.29526	&0.002\\
1.9	&0.30995	&0.30995	&0.001\\
2	&0.32465	&0.32465	&0.000\\  

        	\end{tabular}
        \end{center}
        If the number of discretization points for $x$ is large
        enough, the deviation will become arbitrarily small. Due to
        the simplistic discretization method used by the SciPy ODE
        solver based code, the FlowPy results are likely to be closer to the
        exact solution here.
       
       Now that we have convinced ourselves that {\FlowPy} solves
       these trivial examples correctly, we turn to typical equations
       that are encountered in the study of functional renormalization
       group equations.

\subsection{System of coupled ordinary differential equations}
\label{logfile}
\noindent
In this section we solve the flow equations for the anharmonic
oscillator. This will give us an example for a system of coupled
flow equations. The flow equation for the effective potential, using the regulator $R_k=(k^2-p^2)\theta(k^2-p^2)$,   
is given by \cite{Gies:2006wv}
\begin{equation}\partial_k V_k(x)=\frac{1}{2\pi}\frac{V_k''(x)}{k^2(k^2+V''_k(x))},\end{equation}
where prime denotes the derivative with respect to $x$. In order to
obtain a system of ordinary differential equations we make a
polynomial approximation for $V_k(x)=\tilde
E_k+\omega_kx^2+\lambda_kx^3$. This yields
\begin{align}
	\frac{d}{dk}\tilde E_k&=\frac{1}{\pi}\left(\frac{k^2}{k^2+\omega_k}-1\right),\;\nonumber\\
	\frac{d}{dk} \omega_k&=-\frac{1}{\pi}\frac{k^2}{(k^2+\omega_k)^2}\lambda_k\;,\\
	\frac{d}{dk} \lambda_k& = \frac{6}{\pi}\frac{k^2}{(k^2+\omega_k)^3}\lambda_k^2\;.\nonumber
\end{align}
Because \FlowPy{} by design assumes flow equations to be partial
differential equations involving one extra parameter beyond the
renormalization scale~$k$, we have to use a small workaround if we
want to solve a system of ordinary differential equations. We have to
insert an artificial $x$ dependence in all couplings and solve this
system of pseudo partial differential equations. The values for
couplings is constant for all  $x\in [x_{\rm{start}+1},x_{\rm{end}-1}]$.
A python script to solve these differential equations is shown in
Listing~\ref{fig:ODE}.

{\scriptsize
\begin{lstlisting}[frame=single,caption=An example for a system of ordinary differential equations,label=fig:ODE]
# anharmonic_oscillator.py
import FlowPy 

# specifying the flow equations:
aho_eqns="""
	d/dk E(k,x)=(1/pi)*(k^2/(k^2+omega(k,x))-1);
	d/dk omega(k,x)=-k^2*lambda(k,x)/(pi*(k^2+omega(k,x))^2);
	d/dk lambda(k,x)=(6/pi)*k^2/(k^2+omega(k,x))^3*lambda(k,x)^2;
	
#specifying the starting conditions:
FLOWSTART E(k,x)=0; 
FLOWSTART omega(k,x)=1;
FLOWSTART lambda(k,x)=10;

# === Parameters ===
pi=3.141592653589793;
"""

# specifying the values for x:
xs=[-2,-1,0,1,2]

logger=FlowPy.make_flow_logger(filename="anharmonic_oscillator.flow")

# solving the flow equation:
fp=FlowPy.flowproblem("anharmonic_oscillator",
                  xs=xs,
                  equations=aho_eqns,
                  log_state=logger,
                  decide_iterate=FlowPy.make_lhs_iterator(loops=0),
                  verbose=3
                  )
fp.flow()
\end{lstlisting}}

\newpage

\noindent
The results for $E_k$ obtained with \FlowPy{} for different values of
$\lambda_\Lambda$ are displayed in the table below and coincide to the
third digit with those obtained for example with the
\lstinline!DSolve! routine from \textsc{Mathematica} 5. This deviation
is mostly due to the different discretizations in $k$.
\begin{center}
\begin{tabular}{c|c|c}
	$\lambda_\Lambda$&FlowPy&Mathematica\\\hline
	0&0.49968 &0.49998	\\
	0.1&0.50276&0.50306   	\\
	0.2&0.505756&0.506075  \\
	0.3&0.508676&0.508994\\
	0.4&0.511524 &0.511842 \\
	0.5&0.514306&0.514624\\
	1&0.527365&0.527683 \\
	1.5&0.53928 &0.539598\\
	2&0.550305&0.550623 \\
	2.5&0.56061&0.560928\\
	3&0.570315&0.570634 \\
	4&0.588268&0.588586\\
	5&0.604666&0.604985\\
	6&0.619835&0.620154\\
	7&0.633999&0.634317\\
	8&0.647321&0.647639\\
	9&0.659924&0.660242 \\
	10&0.671905&0.672223	
\end{tabular}

\end{center}

\subsection{Field dependent flow equations}
\begin{figure}
	\centering{
	\includegraphics[width=.6\textwidth]{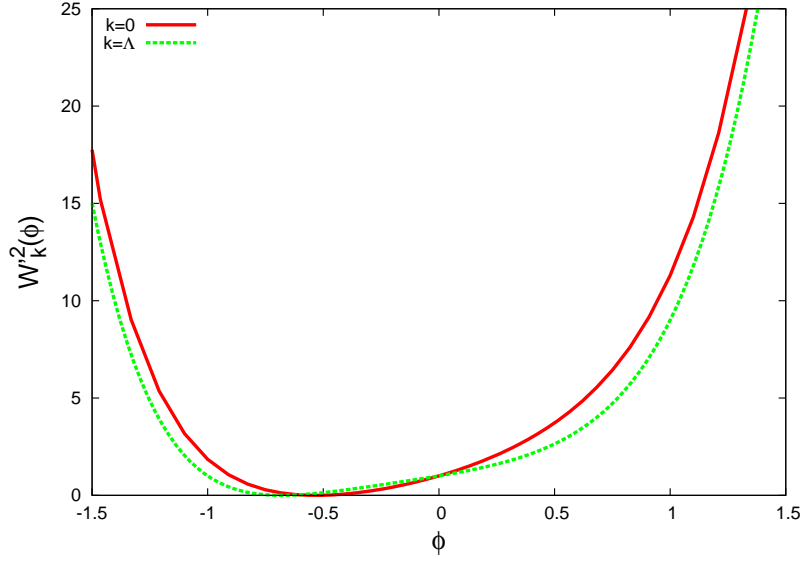}
	\caption{The scalar potential $V=\frac12W'(\phi)^2$ before and after the flow \label{fig:SuSyQM}}}
\end{figure}

\noindent
As an example of a field dependent flow equation we consider
supersymmetric quantum mechanics. This model describes an anharmonic
oscillator coupled in a supersymmetric way to fermions.  The flow
equation is derived for the superpotential $W(\phi)$ which enters in
the scalar potential as $V=\frac12 W'(\phi)$.  The flow equation for
the superpotential reads \cite{Synatschke:2008pv}:
\begin{align}
\partial_k W'_k(\phi) =\frac14\cdot\frac{W_k^{(3)}}{(k+W_k''(\phi))^2}\,.\label{reflow3}
\end{align}
A python script to solve this equation is given in Listing~\ref{fig:FieldDependent}.
{\scriptsize
\begin{lstlisting}[frame=single,caption=An example for a field dependent flow equation,label=fig:FieldDependent]
# SUSY_QM.py
import FlowPy
    
SUSY_QM_eqns="""
	d/dk V(k,phi)=-V''(k,phi)/4/(V'(k,phi)+k)^2;
FLOWSTART V(k,phi) = 1+m*phi+g*phi^2+a*phi^3;

# === Parameters ===
m=1;
g=0;
a=1;
"""

#change n_start and n_end to change the phi range
n_start=-2  
n_end=2
n_steps=10 #increase the number of sampling points to improve the resolution
    
xs=FlowPy.linrange(n_start,n_end,n_steps)

logger=FlowPy.make_flow_logger(filename="SUSY_QM.flow")
fp=FlowPy.flowproblem("flowpy_SUSY_QM",
                  xs=xs,
                  equations=SUSY_QM_eqns,
                  log_state=logger,
                  decide_iterate=FlowPy.make_lhs_iterator(loops=0),
                  verbose=2,
                  )

fp.flow()
\end{lstlisting}}
\noindent
The solution to the flow equation is shown in Fig.~\ref{fig:SuSyQM}. Note that this picture was done with a resolution of \lstinline!n_step=40!. The results are consistent with those obtained with the \lstinline!NDSolve! routine with \textsc{Mathematica} 7 in \cite{Synatschke:2008pv}.

\subsection{System of two coupled field dependent flow  equations}
\noindent
As an example for a system of two coupled flow equations we consider
supersymmetric quantum mechanics with field dependent wave function
renormalization.

\begin{figure}
	\includegraphics[width=.47\textwidth]{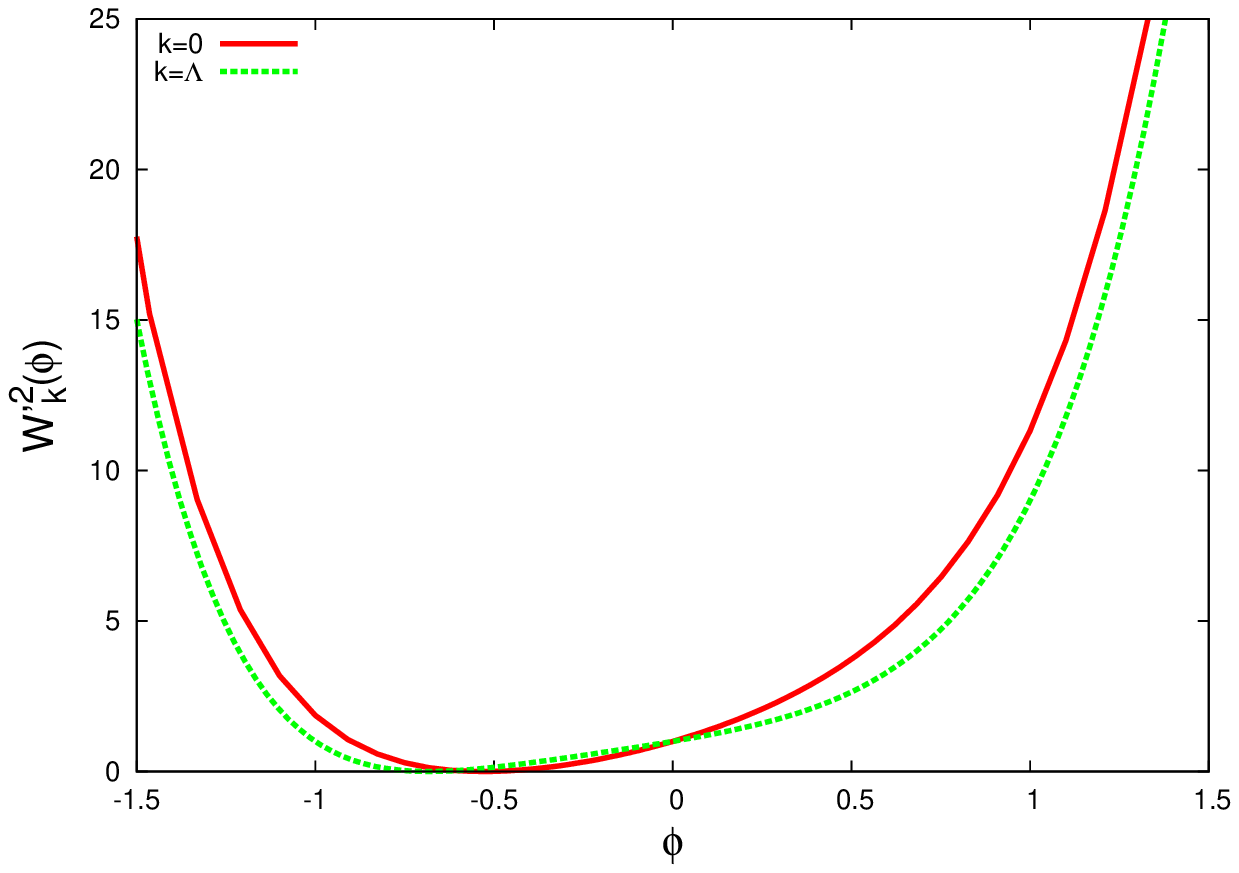}
	\includegraphics[width=.47\textwidth]{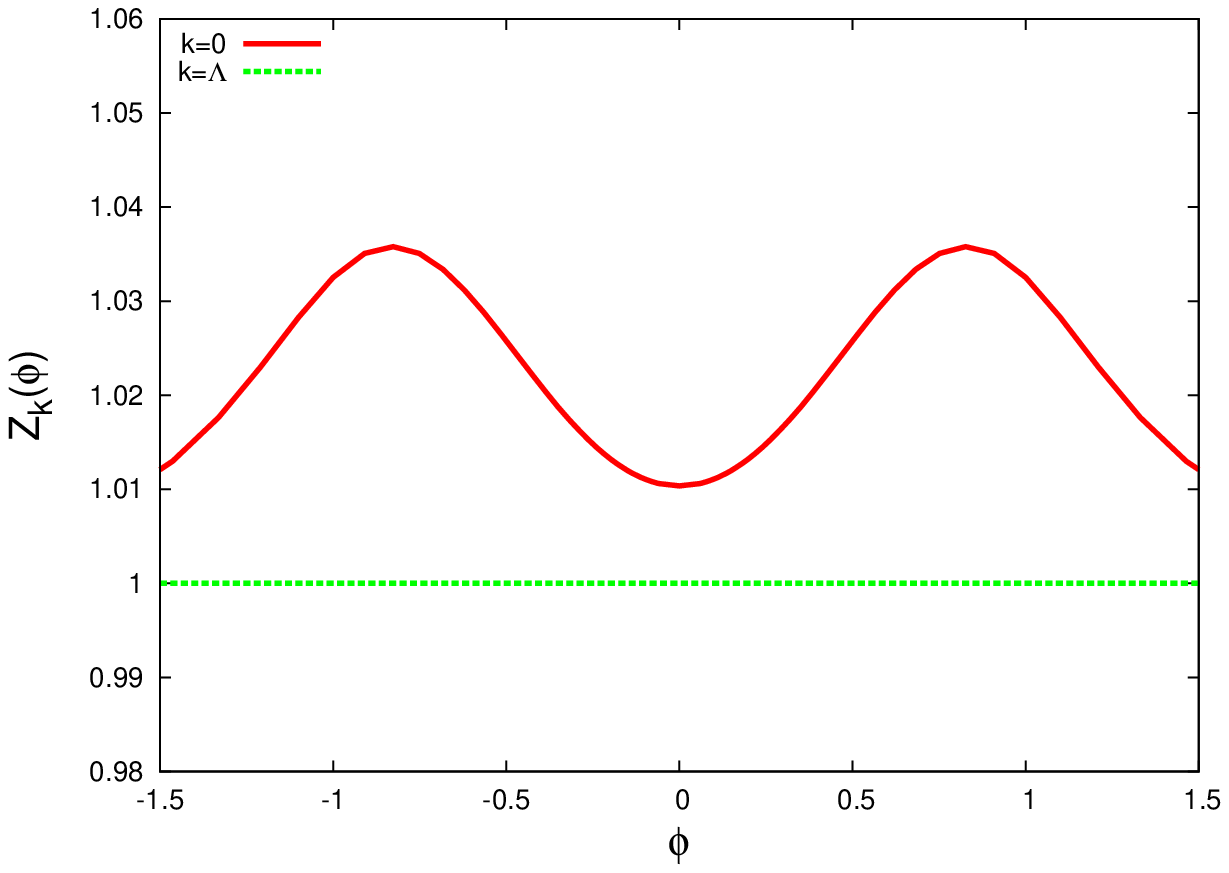}
	\caption{\emph{left panel:} Scalar potential $V=\frac12 W'^2$, \emph{right panel:} field dependent wave function \label{fig:SuSyQMWaveFunction}}
\end{figure}

As in the previous section, the flow equations with a field dependent
wave function renormalization are discussed in
\cite{Synatschke:2008pv}. They read
\begin{equation}
  \begin{aligned}
\partial_k W'_k(\phi)
=& -W'''_k \frac{\N}{4\cD^2}\\
Z_k(\phi)\partial_kZ_k(\phi)=
&\left(
\frac{4Z'_k(\phi)W'''_k(\phi)}{\cD}-\big(Z_k'(\phi)Z_k(\phi)\big)'
-\frac{3Z_k(\phi)^2W'''_k(\phi)^2}{4\cD^2}\right)\frac{\N}{4{\cD}^2}\,,
\label{wfr4}
\end{aligned}
\end{equation}
where we have introduced the abbreviations
\begin{align}
 \N=(1+k\partial_k)Z_k(\phi)^2\quad\text{and}\quad
\cD=W''(\phi)+kZ_k(\phi)^2\,.
\label{wfr5}
\end{align}
The python script to solve these equations is given in
Listing~\ref{fig:CoupledDE}. The solutions are shown in
Fig~\ref{fig:SuSyQMWaveFunction}. This picture was created with \lstinline!n_steps=40!.

{\scriptsize
\begin{lstlisting}[frame=single,caption=An example for a system of two coupled differential equations,label=fig:CoupledDE]
# SUSY_QM_wavefunction.py

import FlowPy 

SUSY_eqns="""
d/dk V(k,phi)=-V''(k,phi)*(Z(k,phi)^2+2*k*Z(k,phi)*d/dk Z(k,phi))
		/4/(V'(k,phi)+k*Z(k,phi)^2)^2;
d/dk Z(k,phi)=(Z(k,phi)+2*k*d/dk Z(k,phi))
		/4/(V'(k,phi)+k*Z(k,phi)^2)^2
		*(4*Z'(k,phi)*V''(k,phi)/(V'(k,phi)+k*Z(k,phi)^2)
			-Z''(k,phi)*Z(k,phi)-Z'(k,phi)*Z'(k,phi)
			-3*Z(k,phi)^2*V''(k,phi)^2/4/(V'(k,phi)
				+k*Z(k,phi)^2)^2);

FLOWSTART V(k,phi) = e+m*phi+g*phi^2+a*phi^3;
FLOWSTART Z(k,phi) = 1;

#==== Parameters ====
e=1.0;
m=1.0;
g=0.1;
a=1.0;
"""
     
#change n_start and n_end to change the phi_range
n_start=-2  
n_end=2
n_steps=10 #increase this variable to improve the resolution

xs=FlowPy.linrange(n_start,n_end,n_steps)

logger=FlowPy.make_flow_logger(filename="SUSY_QM_wavefunction.flow")

fp=FlowPy.flowproblem("flowpy_SUSY_QM_wavefunction",
                  xs=xs,
                  equations=SUSY_eqns,
                  log_state=logger,
                  decide_iterate=FlowPy.make_lhs_iterator(loops=0),
                  verbose=2,
                  )
fp.flow()
\end{lstlisting}}

\subsection{Momentum-dependent flow equations}
\noindent
FlowPy can also solve flow equations with full momentum dependence.
In this example we calculate the momentum dependent wave-function
renormalization for the supersymmetric $N=2$ Wess-Zumino Model in two
dimensions. This model is thoroughly discussed in
\cite{Synatschke:2010jn} where the flow equation is derived. The flow
equation read
\begin{equation}
  \partial_kZ_k^2(p)=-16 g^2\int\frac{d^2 q}{4\pi^2}\frac{k
Z^2_k\left( q\right)+\mbar }
{N(\vect q)^2N(\vect p-\vect q)}
Z^2_k\left(q\right) Z^2_k\left(\abs{\vect p -\vect q}\right)\partial_k
\left(kZ_k^2\left( q\right)\right)   \label{eq:FlowEq},
\end{equation}
where we have introduced the abbreviation
\begin{equation}
N(\vect q)=\left(q^2 Z^4_k\left( q\right)+(kZ^2_k\left(
q\right)+\mbar)^2\right).
\end{equation} 
$\vect q$ and $\vect p$ denote two dimensional vectors, whereas $q=\abs{\vect{q}}$ and $p=\abs{\vect p}$  respectively.

\begin{figure}
\centering{
	\includegraphics[width=.6\textwidth]{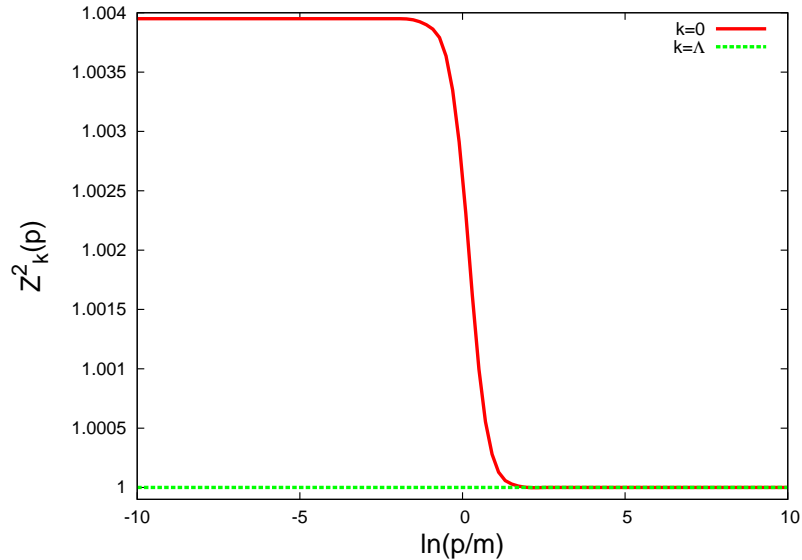}
	\caption{Momentum dependent wave function renormalization \label{fig:WessZuminoWaveFunction}}}
\end{figure}

A python script to solve this equation with $m=1$ and $g=0.3$ is given
in Listing~\ref{fig:DemoWaveFunction}. Note that due to the complex structure of the flow equation this script takes a long time to calculate the wave function even with a very low resolution of five sampling points.  The result is shown in
Fig.~\ref{fig:WessZuminoWaveFunction}. The calculation was done with \lstinline!n_steps=30!.

\newpage

{\scriptsize
\begin{lstlisting}[frame=single,caption=An example for a momentum dependent wave function,label=fig:DemoWaveFunction]
# momentum_dependent_wavefunction.py
import FlowPy
import math

wavefunction_eqns="""
d/dk Z(k,p) =
 integral[d phi from -pi to pi,
          d q from 0 to k]
          (-16*g^2)/4/pi^2*(k*Z(k,q)+m)*Z(k,q)*Z(k,Cminus(p,q,phi))
          *(1+k*d/dk Z(k,q))/(q*q*Z(k,q)*Z(k,q)+(k*Z(k,q)+m)*(k*Z(k,q)+m))
          /(Cminus(p,q,phi)*Cminus(p,q,phi)*Z(k,Cminus(p,q,phi))
          *Z(k,Cminus(p,q,phi))+(k*Z(k,Cminus(p,q,phi))+m)
          *(k*Z(k,Cminus(p,q,phi))+m))
          /(q*q*Z(k,q)*Z(k,q)+(k*Z(k,q)+m)*(k*Z(k,q)+m))^2;
   
FLOWSTART Z(k,p) =1;

# === Helper functions ===

Cminus(p1,p2,alpha) = p1^2+p2^2-2*p1*p2*cos(alpha);

# === Parameters ===
m = 1.0;
g=0.3;
pi=3.141592653589793;
"""

#change n_start and n_end to change the p_range
n_start=-20  
n_end=10
n_steps=5 #increase this number of sampling points to improve the resolution

# logarithmic-equidistant distribution of the momentum:
qs_log=FlowPy.linrange(n_start,n_end,n_steps)
qs=[-math.exp(q) for q in qs_log]+[0.0]+[math.exp(q) for q in qs_log]

# specifing the logfile:
logger=FlowPy.make_flow_logger(filename="momentum_dependent_wavefunction.flow") 

# solving the flow equation:
fp=FlowPy.flowproblem("flowpy_wavefunction",
                  xs=qs,
                  equations=wavefunction_eqns,
                  log_state=logger,
                  decide_iterate=FlowPy.make_lhs_iterator(loops=0),
                  verbose=2)

fp.flow()
\end{lstlisting}}

\subsection{Parametric Example}
\noindent
In some situations, one would want to perform a number of
renormalization group flow calculations that differ only 
in the choice of some parameters. While this can be achieved
in a reasonably straightforward way with {\FlowPy} via
Python scripting, the {\FlowPy} package contains a few extra
definitions to make this process more convenient for the user,
which are explained by means of an example in 
Listing~\ref{fig:parametric}.

{\scriptsize
\begin{lstlisting}[frame=single,caption=Example for a flow equation with two parameters,label=fig:parametric]
# parametric.py
from FlowPy import flowproblem, flowparams, make_lhs_iterator,\
     grange, make_flow_logger

runs_todo=[flowparams(mu=val_mu,T=val_T)
             for val_mu in [1.0,1.5,2.0,2.5]
             for val_T in [0.1,0.2,0.5,1.0]]

xs=[float(n) for n in range(11)]

for params in runs_todo:
    f=flowproblem("mu_T",
                  xs=xs,
                  equations="""
                  d/dk F(k,p) = -T;
                  FLOWSTART F(k,p) = mu;
                  """+params.defs(),
                  ks=grange(110,10,5),
                  decide_iterate=make_lhs_iterator(loops=0),
                  log_state=make_flow_logger(\
                      params.subs_short("FLOW__mu=${mu}_T=${T}__.flow")),
                  verbose=2)
    f.flow()

\end{lstlisting}}
\noindent
The key concept here is the \texttt{flowparams} ``{\FlowPy}
parameters'' object. This represents a collection of parameter 
choices that can easily be mapped to strings in various contexts,
e.\,g. to generate systematic output filenames. A \texttt{flowparams}
object is created by specifying explicitly all ``parameter name =
numerical value'' associations in a function call of the form

{\scriptsize
\begin{lstlisting}
params = flowparams(mu=1.5, alpha=3.0, T=0.5)
\end{lstlisting}}
\noindent
A potentially relevant limitation is that special keywords in the
Python programming language (such as e.\,g. \texttt{def},
\texttt{raise}, \texttt{lambda}, \texttt{else}, etc.) cannot be
chosen as parameter names. (There is a way to avoid this problem in
Python by using a different function call form, should this really
turn out to be a problem.)

If {\lstinline!params!} is a {\FlowPy} parameter object, e.\,g. defined
as {in the example in Listings~\ref{fig:parametric}}, then
{\lstinline!params.defs()!} will produce a string containing valid
{\FlowPy} constant definitions that introduce these parameters. One
would typically want to append this to the (parameter-dependent) flow
equation definitions by using Python ``string addition'' as shown in
the example. Another useful snippet is
{\lstinline!params.subs_short(pattern)!}  which will regard
\texttt{pattern} as a template string on which parameter substitution
has to be performed according to the rules of Python's
\texttt{string.Template} class, i.e. \texttt{\$\{xyz\}} will be
substituted by the pretty-printed numerical value of the parameter
\texttt{xyz}. (Detailed rules can be found in the Python documentation
of the \texttt{string.Template} mechanism.) One may want to use this
to automatically generate filenames, as shown in the example.

A very useful Python feature one might want to employ in such
situations is special syntax available to define lists from cartesian
products. In Python, one can e.\,g. write

{\scriptsize
\begin{lstlisting}
 [(x,y) for x in range(1,5) for y in range(1,x)]
\end{lstlisting}}
\noindent
to obtain the list
{\scriptsize
\begin{lstlisting}
[(2, 1), (3, 1), (3, 2), (4, 1), (4, 2), (4, 3)]
\end{lstlisting}}
\noindent
This syntax, which is an adoption of a very similar feature that
probably was first made popular through the Haskell programming
language, can be used to concisely specify fairly complex
constructions, such as

{\scriptsize
\begin{lstlisting}
from fractions import gcd
[(x,y,x*y) for x in range(5) for y in range(5) if gcd(x,y)==1]
\end{lstlisting}}
\noindent
which gives the list

{\scriptsize
\begin{lstlisting}
[(0, 1, 0),
 (1, 0, 0),
 (1, 1, 1),
 (1, 2, 2),
 (1, 3, 3),
 (1, 4, 4),
 (2, 1, 2),
 (2, 3, 6),
 (3, 1, 3),
 (3, 2, 6),
 (3, 4, 12),
 (4, 1, 4),
 (4, 3, 12)]
\end{lstlisting}}
\noindent
Users of {\FlowPy} might find the basic form of this construct useful
to define cartesian products of parameter choices, as in the example
provided in this section.

Evidently, if the approach shown here is used to perform~$N$ different
numerical renormalization group flow calculations, then~$N$ different
machine code files will be produced automatically by {\FlowPy} and
loaded into Python. Although this process could be improved
conceptually, this would presumably be only worthwhile for
$N\gg100$. This is unlikely to be relevant for typical applications,
as problems of typical complexity are expected to be partitioned into
collections of $N<100$ before being submitted to a computing cluster
anyway. When performing multiple {\FlowPy} calculations on a cluster,
one should ensure that each cluster job chooses a different name when
defining a flowproblem. Otherwise, this may result in accidental unintended sharing of
the directory used to produce machine code, and some
calculations may fail or, even worse, end up using a wrong set of
parameters and equations. Typically, an unique job~ID will be taken
from the program's argument list, from environment variables, or
generated semi-randomly using the time and process ID.  Python makes
this information available via the \texttt{sys.argv} variable
containing program arguments (the \texttt{sys} module has to be
imported first), the \texttt{os.getenv()} function that behaves like
the corresponding C library function, the \texttt{time.time()} and
related functions, and the \texttt{random} module.

\section{Current limitations}
\noindent
In its present form, the {\FlowPy} package has a number of
limitations. None of these are actually fundamental -- they all could
be overcome with some dedicated effort on the side of the FlowPy
authors. Concerning the question how relevant these are, and where to
focus further development effort on, the authors
seek input from the renormalization group flow community.

\noindent
Present  limitations are:
\begin{itemize}
\item {\FlowPy} assumes every flow function to depend on precisely one
  extra parameter beyond the scale parameter~$k$.
\item The right-hand side of flow functions may involve zero to two
  integrals, and the integration boundaries may only depend on
  left-hand side parameters, i.\,e. the inner integration boundary
  cannot depend on the value of the outer integration parameter. Also,
  term notation presently requires integral specifications to 
  follow the equals sign directly, i.\,e. all extra factors must be pulled under
  the integral. Expressions such as $C\int_{\alpha=-\pi}^\pi
  d\alpha\,f(\alpha)$ (with $C$~a numerical constant) hence must be
  re-written as $\int_{\alpha=-\pi}^\pi d\alpha\,Cf(\alpha)$.

\item{Currently boundary conditions in the $x$ direction are fixed to
    the classical values of the effective potential. However, more
    flexible boundary conditions that can be specified by the user
    might be more appropriate for some problems.}

\item The range of special functions made available to the user so far
  is fairly limited.

\item Error reporting could be improved to make it easier to find
  typos in the specification of the flow equations.

\item {\FlowPy} suffers from some minor bugs in the default parser
  definitions inherited from \texttt{python-simpleparse}. In
  particular, a number such as `\texttt{5e-3}' is not a valid constant
  -- this must be given as `\texttt{5.0e-3}' instead.

\item This version of FlowPy does not support
  parallelization. However, re-introducing MPI support is planned.

\item There are many opportunities to make {\FlowPy} more efficient by
  improving its internal design. This, however, will require
  substantial low level changes.

\end{itemize}

\section{Outlook}
\noindent
In this paper we present the numerical toolbox \FlowPy{} which is able
to solve many typical partial differential equations encountered in
studies of the functional renormalization group. We hope that it will
prove to be useful in the application of functional renormalization
group techniques and that it will facilitate these studies in
providing a powerful numerical tool to solve the differential
equations.

We plan to include MPI support in the next version of
\FlowPy{}. Additionally it is planned to make the boundary conditions
in the field variable more flexible such that it will be possible to
use e.\,g. the first loop approximation. Also there is a lot of room for
improvement in the performance of the parser. Here we present the
first version of \FlowPy{} and we have planned to release updates in
the future. The intention of this paper is to make \FlowPy{} available
to a wider audience. Thus we are grateful for suggestions how to improve
\FlowPy{}.

\section{Acknowledgments}
The authors would like to thank J.~Braun, H.~Gies and
A.~Wipf for helpful discussions and valuable comments on the
manuscript. Also, the authors would like to thank Max Albert
for helping with the effort to port {\FlowPy} to the Mac OS X
platform. Part of this work was supported by the German Science
Foundation (DFG) under GRK 1523 and the Studienstiftung des
deutschen Volkes e.\,V.



\end{document}